\documentclass{article}
\usepackage{spconf,amsmath,graphicx}

\usepackage{multirow} 
\usepackage{ulem}
\useunder{\uline}{\ul}{}
\usepackage{amsfonts}

\title{TEACHER-STUDENT NETWORK FOR REAL-WORLD FACE SUPER-RESOLUTION WITH PROGRESSIVE EMBEDDING OF EDGE INFORMATION}
\name{Zhilei Liu\sthanks{Corresponding author, Email: zhileiliu@tju.edu.cn}, Chenggong Zhang}
\address{College of Intelligence and Computing, Tianjin University, Tianjin, China}

\begin{document}
%\ninept
%
\normalem
\maketitle

\begin{abstract}
Traditional face super-resolution (FSR) methods trained on synthetic datasets usually have poor generalization ability for real-world face images. Recent work has utilized complex degradation models or training networks to simulate the real degradation process, but this limits the performance of these methods due to the domain differences that still exist between the generated low-resolution images and the real low-resolution images. Moreover, because of the existence of a domain gap, the semantic feature information of the target domain may be affected when synthetic data and real data are utilized to train super-resolution models simultaneously. In this study, a real-world face super-resolution teacher-student model is proposed, which considers the domain gap between real and synthetic data and progressively includes diverse edge information by using the recurrent network's intermediate outputs. Extensive experiments demonstrate that our proposed approach surpasses state-of-the-art methods in obtaining high-quality face images for real-world FSR.
\end{abstract}

\vspace{-4mm}
\section{Introduction}
\label{sec:intro}
\vspace{-2mm}
Face super-resolution (FSR), also known as face hallucination, aims to reconstruct high-resolution (HR) face images from their degraded low-resolution (LR) counterparts. It is difficult for FSR to construct a super-resolved image from a very low-resolution image (eg. 16$\times$16) since facial components are distorted. To promote the performance of FSR, recent works~\cite{ma2020deep,kim2021progressive,zhang2022face} integrated facial landmarks and component maps into the FSR problem to enhance FSR performance. Despite the fact that certain techniques, such as bicubic interpolation downsampling, have produced acceptable results on LR photos, the degradation of low-resolution images in real-world scenarios is typically more difficult. These SR networks trained on synthetic datasets usually lead to undesired strong artifacts in their SR results when applied to real images.  
%these methods are trained on synthetic datasets and their poor generalization capacity limits their application in real scenarios.
As a result, attempts to solve the real-world SR problem have been made during the past few years~\cite{bulat2017far,wang2021real,wei2021unsupervised,hou2023semi}.  Specifically, Bulat et al.~\cite{bulat2017far} proposed LRGAN which uses a "high-to-low" branch network to learn the real degradation process of real-world LR images. The methods of GLEAN~\cite{chan2021glean} and GFPGAN~\cite{wang2021towards} achieve promising performance for FSR with implicit priors encapsulated in a pre-trained GAN, in which complicated degradation models or training networks are investigated to simulate the actual deterioration process. Unfortunately, the performances of these approaches are still constrained by the domain gap between synthetic LR images and real LR images~\cite{wei2021unsupervised}. To tackle the real-world image SR problem, Wei et al.~\cite{wei2021unsupervised} proposed a domain-distance aware super-resolution approach that assigns different weights to various data based on the domain distance. Hou et al.~\cite{hou2023semi} used three independent
branches to learn the forward and backward cycle-consistent reconstruction processes to  alleviate the domain gap
between unpaired LR and HR face images. However, when synthetic data and real data are used to train super-resolution models at the same time, the semantic feature information of the target domain may be damaged due to the existence of the domain gap. Moreover, most of the existing real-world FSR methods usually ignore facial prior information.

To solve these problems mentioned above, we propose a teacher-student network for real-world face super-resolution with progressive edge information embedding. In accordance with prior work~\cite{bulat2018learn}, we synthesize pseudo-paired data using a degraded network to discover the real degradation process, and then employ these data to train a teacher super-resolution network (TNet). Because of the domain gap between the synthetic data and the real data, we only use real low-resolution images to train a student super-resolution network(SNet). Existing approaches often use the cycle learning scheme to tackle super-resolution tasks when there are no paired data~\cite{zhu2017unpaired}. However, because only adversarial loss is supported in high-resolution space, the generated face image frequently suffers structural deformation. Therefore, we use the teacher-student framework to generate pseudo high-resolution images for real low-resolution images to assist in the reconstruction of face structure and maintain pixel-level accuracy in the student network. The low-frequency and high-frequency components of a single picture may usually be identified intuitively. The goal of the SR task is to restore the high-frequency component of the image, which is essential for constructing the object's structure. In this paper, we propose to use the edge information of the face to enhance the restoration of the global shape and local details in the reconstruction process of the face image combined with the traditional image processing method. Moreover, to improve the recovery of edge information, we employ a recurrent convolutional network as the SR backbone and progressively embed various edge information at different stages.

\vspace{-4mm}
\section{PROPOSED METHOD}
\label{sec:majhead}
\vspace{-2mm}

\subsection{Overviews}
\vspace{-2mm}

\begin{figure}[h]
\vspace{-1mm}
\begin{center}
%\fbox{\rule{0pt}{2in} \rule{.9\linewidth}{0pt}}
 \includegraphics[width=\linewidth]{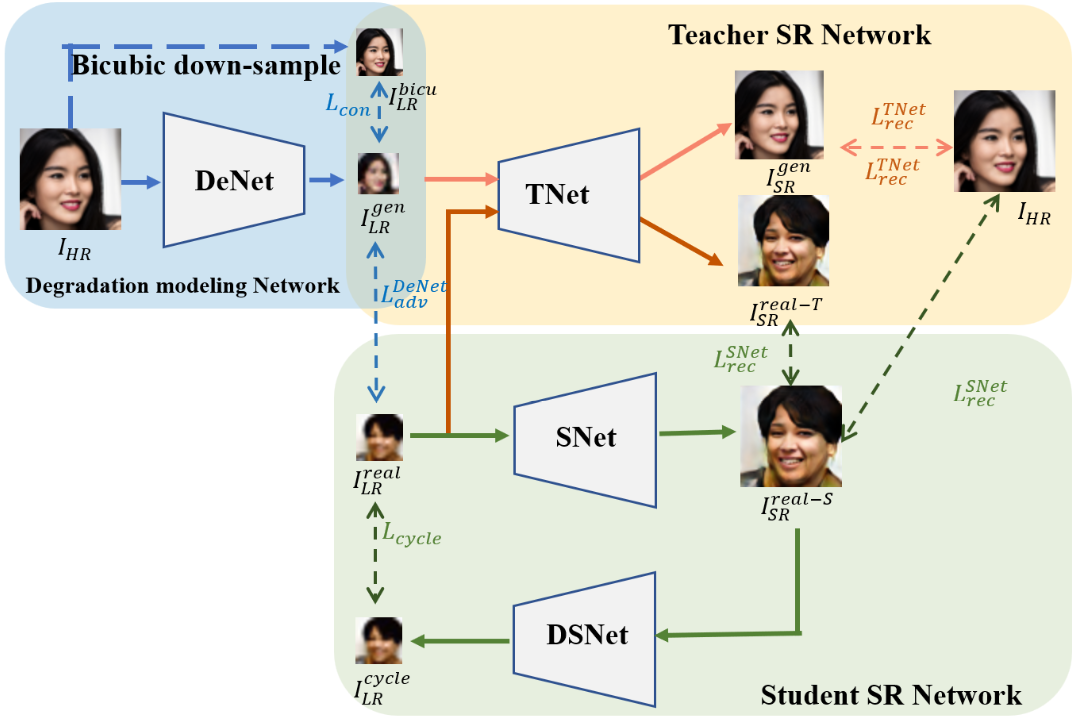}
\end{center}

\vspace{-6mm}
   \caption{Pipeline of our proposed method.}
\label{fig:Pipeline}
\vspace{-2mm}
\end{figure}
\vspace{-2mm}

For real-world FSR, the data that can be obtained are usually two sets of unpaired LR images $I_{LR}$ and HR images $I_{HR}$. To achieve real-world FSR, a two-stage framework following DASR~\cite{wei2021unsupervised} is proposed, and the overall pipeline is shown in Fig.~\ref{fig:Pipeline}. First, a degradation modeling Network (DeNet) is trained to generate an LR image $I_{LR}^{gen}$ that conforms to the real LR domain distribution as much as possible. Then, the generated LR-HR image pairs are used to train our Teacher super-resolution Network (TNet). Different from the previous method, we fully consider the domain gap between the synthetic LR image and the real LR image and design a Student Network (SNet) to be specially used for the super-resolution reconstruction of the real LR domain image $I_{LR}^{real}$. Because there are no HR images corresponding to the real LR image, we use the $I_{SR}^{real-T}$ obtained by TNet to help SNet maintain facial features and pixel-level accuracy during training. Using the data synthesized by DeNet, the results of TNet trained in a supervised manner have a better recovery of face features, so it can be used to promote the convergence of SNet during the training phase. In addition, we also use a Down-Sampling Network (DSNet) to downsample $I_{SR}^{real-S}$ to the LR domain to use cycle consistency constraints in the LR domain to better train the SNet. 
%Because there is no reliable reference image for the LR image in the super-resolution task in the real scene, it cannot completely rely on the reconstruction loss training. 

The adversarial loss can make the generated image more realistic perceptually, so all the networks are trained in an adversarial learning manner. DeNet, TNet, SNet, and DSNet are the generators of each generative adversarial network, and the structure of the discriminator adopts LightCNN~\cite{wu2018light}. For ease of representation, all discriminator parts are omitted in Fig.~\ref{fig:Pipeline}. We retain SNet around there for experimental network verification during testing.

\vspace{-4mm}
\subsection{Network Architecture}
\vspace{-2mm}

\begin{figure}[h]
\vspace{-2mm}
\begin{center}
%\fbox{\rule{0pt}{2in} \rule{.9\linewidth}{0pt}}
 \includegraphics[width=\linewidth]{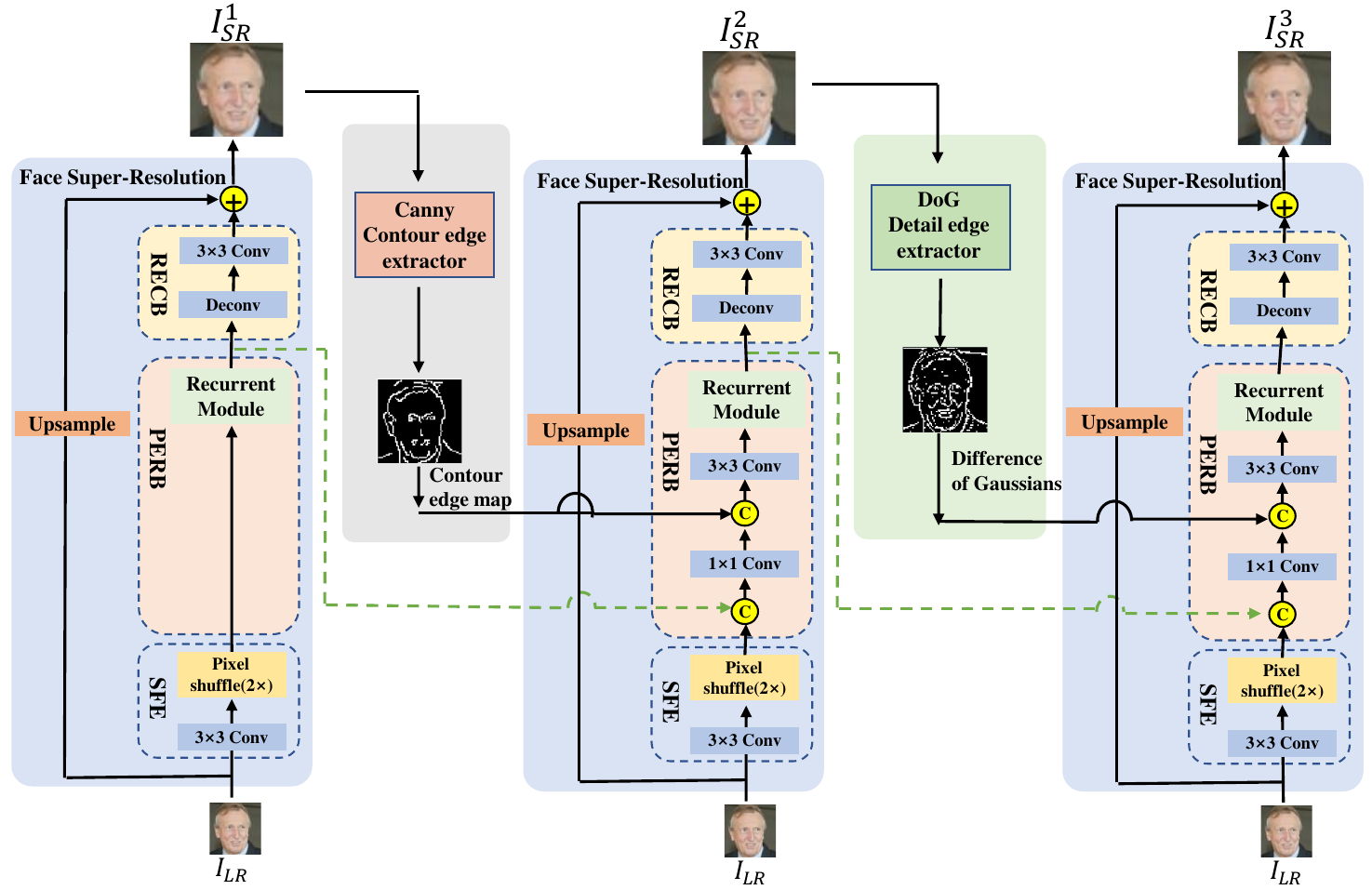}
\end{center}

\vspace{-6mm}
   \caption{The unfolded architecture of TNet and SNet.“C”,"+", and “×” denote concatenation, addition, and multiplication, respectively. The green dotted lines represent feedback connections. For simplicity, we omit the activation function layer in the pipeline.}
\label{fig:TNet}
\vspace{-2mm}
\end{figure}
\vspace{-2mm}

\textbf{DeNet and DSNet}: DeNet takes the HR image as input, uses a convolutional layer and 16 residual blocks~\cite{he2016deep} to extract features in high-resolution space, and then uses a downsampling layer containing two convolutional layers with a stride of 2 to reduce the spatial resolution of the features to the required size, and finally, the features are mapped back to the image domain through a convolutional layer. The network structure of DSNet is the same as that of DeNet.

\textbf{SR Network}: As shown in Figure~\ref{fig:TNet}, our proposed TNet and SNet can be unfolded into 3 iterations. The face super-resolution branch in each iteration contains three parts: a shallow feature extractor (SFE), a prior embedded recurrent block (PERB), and a high-resolution reconstruction block (RECB). Given a low-resolution (LR) input $ I_{LR} $, we use a 3×3 convolutional layer and a pixel shuffle layer to extract shallow feature $ F^t_{sf} $ at $ t$-th iteration.
% as:
% \begin{equation}
% \label{eq_sf}
%  F^t_{sf} = H_{SFE}(I_{LR}),
% \end{equation}
% where $H_{SFE}(\cdot)$ denotes the operations of the SFE. 
Then, we use a 1×1 and a 3×3 convolutional layer to fuse shallow feature $ F^t_{sf} $, the feedback feature $ F^{t-1}_{fb} $ from the previous iteration and prior information $ F_{prior} $, and use a recurrent module to generate high-level representations $ F^{t}_{fb}$. Specifically, the architecture of the recurrent module follows the feedback block in~\cite{li2019feedback} and we remove its first convolutional layer. Note that there is no prior information and hidden state at the first iteration in our model. Then, the outputs of the PERB are used as the input to the RECB to generate an SR residual image of the HR face $ I^t_{Res}$.
% as :
% \begin{equation}
% \label{eq_res}
%  I^t_{Res} = H_{RECB}(F^{t}_{fb}),
% \end{equation}
% where $H_{RECB}(\cdot)$ denotes the operations of the RECB. 
RECB consists of a deconvolutional layer and a 3×3 convolutional layer. Finally, the restored SR image at the $t$-th iteration can be described as:
\begin{equation}
\footnotesize
\label{eq_sr}
 I^t_{SR} = I^t_{Res} + H_{UP}(I_{LR}),
\end{equation}
where $H_{UP}(\cdot)$ denotes an bilinear upsampling operation. Therefore, we will get totally 3 SR images ($I^1_{SR}$, $I^2_{SR}$, $I^3_{SR}$) for every LR image $I_{LR}$.

\textbf{Edge Detector}: LR images in real scenes are often affected by noise, so we choose to use the Canny edge detector that can handle noise well to extract the high-frequency edge part of the overall contour of the image. In addition, the Laplace of Gaussian function (LoG) can obtain the local details of the image, but LoG needs to perform the second-order difference or second-order partial derivative of the discrete digital image, and the calculation amount is relatively large. The Difference of Gaussians (DoG) of different scales can be used as an approximate representation of LoG. Therefore, we choose to use the Canny edge detector to extract the global contour edge of the face image from the output of the first iteration and input it into the next step to splicing with the feature map in the backbone network to promote the restoration of facial structure. And then use the DoG operator to extract the differential detail edge map of the image in the output of the second iteration, and input it into the third iteration to further promote the recovery of high-frequency information of local details of the face. 
%so as to explicitly enable the network to pay more attention to the restoration of high-frequency information of the face improves the reconstruction quality of the face structure and details.
\vspace{-3mm}

\subsection{Loss functions}
\label{ssec:subhead}
\vspace{-2mm}
For the DeNet, in order to make the generated LR image consistent with the content of the HR image, we use the content loss to constrain the content between $I_{LR}^{gen}$ and $I_{LR}^{bic}$ obtained by bicubic downsampling from the HR image, which is defined as follows:
\vspace{-1mm}
\begin{equation}
\footnotesize
\label{context_loss}
L_{con} = \mathbb{E}\big|\big|I_{LR}^{gen} - I_{LR}^{bic}\big|\big|_1 
\end{equation}

To achieve the conversion to the real LR domain, we apply an adversarial loss between $I_{LR}^{real}$ and $I_{LR}^{gen}$, which is defined as follows:

% \begin{equation}
% \label{eq_loss_advg_denet}
% L_{adv\_G}^{DeNet} = - \mathbb{E}\big[log(D_1(I_{LR}^{gen})))\big]
% \end{equation}

% and for  discriminator $D_1$:
\vspace{-4mm}
\begin{equation}
\footnotesize
\label{eq_loss_advd_denet}
L_{adv}^{DeNet} = \mathbb{E}\big[log(D(I_{LR}^{real}))\big] + \mathbb{E}\big[log(1 - D(I_{LR}^{gen})))\big]
\end{equation}

%where $I_{LR}^{real}$ and $I_{HR}$ are unpaired. 
Therefore, the overall loss function of DeNet is:
\vspace{-3mm}
\begin{equation}
\footnotesize
\label{disloss}
L_{DeNet} = \alpha_1 L_{con} + \beta_1 L_{dis_1}
\vspace{-2mm}
\end{equation}

where $\alpha_1$, $\beta_1$ are trade-off parameters.
For TNet, we use LR-HR data pairs synthesized by DeNet to train in a supervised manner. To achieve better optimization, TNet uses reconstruction loss $L_{dis_2}$ as:
\vspace{-2mm}
\begin{equation}
\label{eq_loss_rec}
\footnotesize
L_{rec}^{TNet} = \mathbb{E}\bigg[\frac{1}{N}\sum_{n=1}^{N}||I_{HR}-I^n_{SR}||^2_2 \bigg],
\end{equation}
and adversarial loss $L_{adv}^{TNet}$ is similar with Eq.~\ref{disloss}
% \begin{equation}
% \label{eq_loss_advg}
% L_{adv\_G}^{TNet} = - \mathbb{E}[log(D(G(I_{LR})))],
% \end{equation}
Therefore, the total loss function when training TNet is:

\begin{equation}
\footnotesize
\label{eq_loss_all_tnet}
L_{TNet} = \alpha_2 L_{rec}^{TNet} + \beta_2 L_{L_{adv}^{TNet}}
\end{equation}

For SNet, we fully consider the domain gap and only use the real LR images to train SNet. We use the pseudo HR images generated by TNet as the corresponding real LR image for supervision to guide SNet to better optimize and converge, constrained using a content loss $L_{rec}^{SNet}$ similar with Eq.~\ref{eq_loss_rec}

% \begin{equation}
% \label{eq_loss_rec_snet}
% L_{rec}^{SNet} = \mathbb{E}\bigg[\frac{1}{N}\sum_{n=1}^{N}\big|\big|TNet(I_{LR}^{real})_n-SNet(I_{LR}^{real})_n\big|\big|^2_2 \bigg]
% \end{equation}

In addition, in order to further promote the reconstruction of real LR images, we adopt the idea of CycleGAN~\cite{zhu2017unpaired}, and introduce DSNet to impose cycle consistency constraints on LR images:

\begin{equation}
\footnotesize
\label{eq_loss_all_cyc}
L_{cycle} = \mathbb{E}\big|\big|I_{LR}^{real} - DSNet(SNet(I_{LR}^{real}))\big|\big|_2^2
\end{equation}

To make the generated SR image conform to the distribution of the HR image, we apply the adversarial loss between $I_{SR}^{SNet}$ and $I_{HR}$, the adversarial loss $L_{adv}^{SNet}$ is similar with Eq.~\ref{disloss}. Therefore, the overall loss function of SNet is:

\begin{equation}
\footnotesize
\label{eq_loss_all_snet}
L_{SNet} = \alpha_3 L_{rec}^{SNet} + \beta_3 L_{adv}^{SNet} + \gamma L_{cycle}
\end{equation}

\vspace{-2mm}
\section{EXPERIMENTS}
\label{sec:foot}
\vspace{-2mm}

\subsection{Experimental Setup}
\vspace{-2mm}

Following existing work SCGAN\cite{hou2023semi}, we use the 20,000 high-quality, high-resolution face images from FFHQ dataset\cite{karras2019style} and the 4,000 low-resolution face images from the Widerface\cite{yang2016wider} for training, and use two synthetic datasets, i.e., LS3D-W
balanced\cite{bulat2017far} and FFHQ\cite{karras2019style}, and two real-world datasets,
i.e., Widerface\cite{yang2016wider} and Webface\cite{hou2023semi} for testing. All the above datasets are processed in the same way as SCGAN\cite{hou2023semi}. All experiments were carried out at 4× super-resolution factor.

For quantitative evaluation, the non-reference metric FID~\cite{heusel2017gans} and full reference metrics (LPIPS~\cite{zhang2018unreasonable}, PSNR, and SSIM) are used for two synthetic datasets with corresponding high-resolution reference images. For two real-world datasets, the widely used metrics FID and NIQE~\cite{mittal2012making} are used. The training process contains two stages. First, we separately train the degradation modeling down-sampling network, and set $\alpha_1 = 1$, $\beta_1 = 0.05$, and the batch size is set to 16. Then, we use the trained DeNet to generate its corresponding image from the HR image, which is close to the real LR image domain, for the second stage of training. 
In the second stage, we jointly train TNet and SNet with $\alpha_2 = 1$, $\beta_2 = 0.001$, $\alpha_3 = 1$, $\beta_3 = 0.001$, $\gamma = 1$. The batch size is set to 4. Canny and DoG edges are extracted based on the OpenCV library. The initial learning rate of all networks is set to $1 \times 10^{-4}$, halve every 10 epochs. And we use the ADAM optimizer~\cite{kingma2014adam} for training the model.

\vspace{-4mm}
\subsection{Comparisons with the state-of-the-art methods}
\vspace{-2mm}

\begin{table}[t]
\setlength{\tabcolsep}{1pt}
\scriptsize
%\begin{threeparttable}
\centering 
\caption{Quantitative comparison with state-of-the-art methods on two synthetic datasets. The best and second best results are marked with \textbf{bold} and \uline{underline} respectively, where "*" indicates that the method indicators come from SCGAN~\cite{hou2023semi}, "-" indicates that the corresponding result is not listed in SCGAN, the same below.}
\resizebox{\linewidth}{!}{
\begin{tabular}{c|cccc|cccc}
\hline
\multirow{2}{*}{Methods}                              & \multicolumn{4}{c|}{LS3D-W balanced}                                                                                                  & \multicolumn{4}{c}{FFHQ}                                                                                                             \\ \cline{2-9} 
                                                 & \multicolumn{1}{c|}{FID$\downarrow$} & \multicolumn{1}{c|}{LPIPS$\downarrow$} & \multicolumn{1}{c|}{PSNR$\uparrow$} & SSIM$\uparrow$  & \multicolumn{1}{c|}{FID$\downarrow$} & \multicolumn{1}{c|}{LPIPS$\downarrow$} & \multicolumn{1}{c|}{PSNR$\uparrow$} & SSIM$\uparrow$  \\ \hline
Bicubic                                          & \multicolumn{1}{c|}{271.86}          & \multicolumn{1}{c|}{0.349}             & \multicolumn{1}{c|}{\uline{ 23.25}}    & 0.6723          & \multicolumn{1}{c|}{276.67}          & \multicolumn{1}{c|}{0.377}             & \multicolumn{1}{c|}{20.68}          & 0.6525          \\ \hline
DASR\cite{wei2021unsupervised}  & \multicolumn{1}{c|}{115.84}          & \multicolumn{1}{c|}{0.360}             & \multicolumn{1}{c|}{23.06}          & 0.6594          & \multicolumn{1}{c|}{113.01}          & \multicolumn{1}{c|}{0.398}             & \multicolumn{1}{c|}{20.72}          & 0.6251          \\ \hline
Real-ESRGAN*\cite{wang2021real} & \multicolumn{1}{c|}{57.20}           & \multicolumn{1}{c|}{0.114}             & \multicolumn{1}{c|}{-}              & -               & \multicolumn{1}{c|}{43.75}           & \multicolumn{1}{c|}{0.336}             & \multicolumn{1}{c|}{-}              & -               \\ \hline
LRGAN\cite{bulat2018learn}      & \multicolumn{1}{c|}{25.35}           & \multicolumn{1}{c|}{0.241}             & \multicolumn{1}{c|}{21.70}          & 0.6810          & \multicolumn{1}{c|}{15.30}           & \multicolumn{1}{c|}{0.296}             & \multicolumn{1}{c|}{20.96}          & 0.6496          \\ \hline
GFPGAN*\cite{wang2021towards}   & \multicolumn{1}{c|}{51.02}           & \multicolumn{1}{c|}{0.094}             & \multicolumn{1}{c|}{-}              & -               & \multicolumn{1}{c|}{43.86}           & \multicolumn{1}{c|}{0.299}             & \multicolumn{1}{c|}{-}              & -               \\ \hline
SCGAN\cite{hou2023semi}         & \multicolumn{1}{c|}{\textbf{20.38}}  & \multicolumn{1}{c|}{\textbf{0.068}}    & \multicolumn{1}{c|}{23.17}          & \uline{0.7535}    & \multicolumn{1}{c|}{\textbf{9.11}}   & \multicolumn{1}{c|}{\uline{ 0.205}}       & \multicolumn{1}{c|}{\uline{ 21.56}}    & \uline{0.7197}    \\ \hline
Ours                                             & \multicolumn{1}{c|}{\uline{23.00}}     & \multicolumn{1}{c|}{\uline{0.084}}       & \multicolumn{1}{c|}{\textbf{24.78}} & \textbf{0.7717} & \multicolumn{1}{c|}{\uline{11.89}}     & \multicolumn{1}{c|}{\textbf{0.203}}    & \multicolumn{1}{c|}{\textbf{21.95}} & \textbf{0.7377} \\ \hline
\end{tabular}
}
\label{tab:sys}
%\end{threeparttable}
\vspace{-4mm}
\end{table}

We compare our proposed method with state-of-the-art SR methods, including generic SR methods like DASR\cite{wei2021unsupervised}, Real-ESRGAN\cite{wang2021real}, and face restoration methods like LRGAN\cite{bulat2018learn}, GFPGAN\cite{wang2021towards}, SCGAN\cite{hou2023semi}. Bicubic interpolation is also introduced as a baseline. For a fair comparison, we use the same data as in SCGAN when training the models. 
%LRGAN and DASR are trained on the same dataset using their published codes. SCGAN uses the pre-trained model provided by it for testing, all quantitative results are obtained with the same calculation code, and the results of other methods refer to the results in the SCGAN\cite{hou2023semi}.

The quantitative results of different methods on the synthetic datasets are shown in Table~\ref{tab:sys}. It can be observed that our method achieves the highest PSNR and SSIM on two datasets, and achieves the lowest LPIPS value on the FFHQ dataset. In addition, our FID on the two datasets also achieves the second-best and the competitive results with the method SCGAN~\cite{hou2023semi}, which shows that the results of our method can be well close to the distribution of real high-resolution face images and natural images while maintaining the pixel-level structure of the reference high-resolution images. 
The quantitative results of different methods on the synthetic datasets are shown in Table~\ref{tab:real}. It can be seen that our method obtains the lowest NIQE value and the second lowest FID value on the two datasets, indicating that the results obtained by our method can be very close to the distribution of real face images and natural images. It is worth noting that those ‘perceptual’ metrics are well correlated with the human-opinion-scores on a coarse scale, but not always well correlated on a finer scale~\cite{wang2021towards}. 

\begin{figure}[h]
\begin{center}
%\fbox{\rule{0pt}{2in} \rule{.9\linewidth}{0pt}}
 \includegraphics[width=0.99\linewidth]{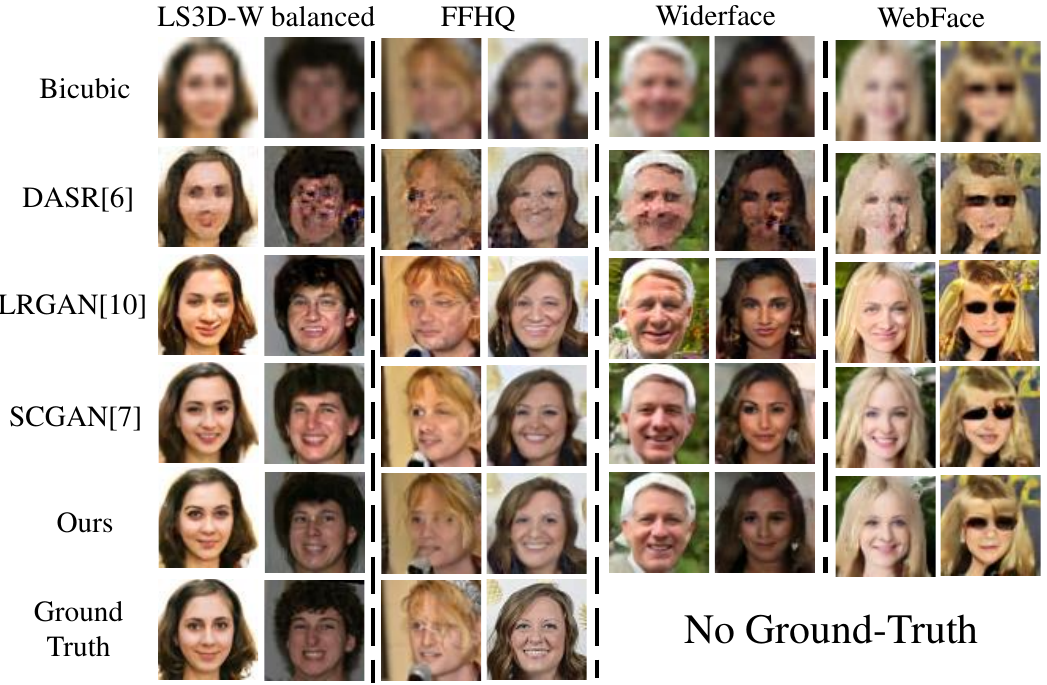}
\end{center}

\vspace{-6mm}
   \caption{Qualitative comparison with state-of-the-art methods on different datasets.}
\label{fig:Qualitative}
\vspace{-6mm}
\end{figure}

Qualitative results on four datasets are shown in Fig.~\ref{fig:Qualitative}. It can be observed that our method is well consistent with the ground truth in terms of expression (such as the first column), skin color (such as the second column), and face structure and details. It also can be observed that, on real-world datasets, our method is more natural in the recovery of face structures (such as contours, eyes, mouth, etc.). In addition, combined with the results of LR images and Bicubic, it can be seen that ours is also better than LRGAN and SCGAN in terms of skin color and overall color restoration, which shows the necessity of our method to fully consider the domain gap.

%It also can be observed that, on real-world datasets, our method is more natural in the recovery of face structures (such as contours, eyes, mouth, etc.), indicating that The method in this chapter introduces the effect of edge information on the restoration of face structure and details; in addition, combined with the results of LR images and Bicubic, it can be seen that the method in this chapter is also better than LRGAN in terms of skin color and overall color restoration. , SCGAN method keeps better, this is because these two methods either only train the network with synthetic data, or train the network with synthetic data and real data at the same time, and the domain gap between the synthetic data and the real data causes the generator to be It is more inclined to learn the distribution of synthetic data, which also shows the necessity of our method to fully consider the domain gap.

% \usepackage{multirow}
% \usepackage[normalem]{ulem}
\begin{table}[]
\tiny
\centering 
\caption{Quantitative comparison with state-of-the-art methods on two real-world datasets.}
\resizebox{\linewidth}{!}{
\begin{tabular}{c|cc|cc}
\hline
\multirow{2}{*}{Methods}                         & \multicolumn{2}{c|}{Widerface}                          & \multicolumn{2}{c}{WebFace}                            \\ \cline{2-5} 
                                                 & \multicolumn{1}{c|}{FID$\downarrow$} & NIQE$\downarrow$ & \multicolumn{1}{c|}{FID$\downarrow$} & NIQE$\downarrow$ \\ \hline
Bicubic                                          & \multicolumn{1}{c|}{269.22}          & 10.8979          & \multicolumn{1}{c|}{273.46}          & 11.2349          \\ \hline
DASR\cite{wei2021unsupervised}                  & \multicolumn{1}{c|}{102.79}          & 8.5417           & \multicolumn{1}{c|}{105.74}          & 8.6894           \\ \hline
Real-ESRGAN*\cite{wang2021real} & \multicolumn{1}{c|}{41.73}           & 6.7595           & \multicolumn{1}{c|}{51.57}           & 6.6773           \\ \hline
LRGAN\cite{bulat2018learn}      & \multicolumn{1}{c|}{16.50}           & 6.9084           & \multicolumn{1}{c|}{23.23}           & 7.0073           \\ \hline
GFPGAN*\cite{wang2021towards}   & \multicolumn{1}{c|}{59.34}           & 6.9437           & \multicolumn{1}{c|}{88.71}           & 6.8841           \\ \hline
SCGAN\cite{hou2023semi}         & \multicolumn{1}{c|}{\textbf{13.05}}  & \uline{6.6192}     & \multicolumn{1}{c|}{\textbf{21.26}}  & \uline{6.5835}     \\ \hline
Ours                                             & \multicolumn{1}{c|}{\uline{14.31}}     & \textbf{6.0863}  & \multicolumn{1}{c|}{\uline{22.49}}     & \textbf{6.1948}  \\ \hline
\end{tabular}
}
\label{tab:real}
\vspace{-4mm}
\end{table}

\vspace{-4mm}

% \begin{figure*}
% \begin{center}
% %\fbox{\rule{0pt}{2in} \rule{.9\linewidth}{0pt}}
%  \includegraphics[width=0.95\linewidth]{Quality.pdf}
% \end{center}
%    \caption{Qualitative comparison with state-of-the-art methods on different datasets.}
% \label{fig:Qualitative}
% \end{figure*}

\vspace{-2mm}
\subsection{Ablation study}
\vspace{-2mm}
To explore the effectiveness of each part of the proposed method, we further implement ablation studies on the FFHQ dataset and the Widerface dataset. 
%All variant models in this part of the ablation study undergo the same number of training steps.
First, the effect of introducing different edge information is analyzed. The baseline model (baseline) in this part means that no contour edge information (expressed in Canny) and Gaussian difference detail edge (expressed in DoG) are added. The quantitative results of the ablation study are shown in the upper part of Table~\ref{tab:ablation}. It can be seen that after gradually adding Canny and DoG, the metrics are gradually getting better.
%, which shows that the introduction of prior information will help the model learn the distribution of real high-resolution face images.
%In addition, it can also be seen that the baseline model obtains higher PSNR and SSIM values than the model after introducing the contour edge (Canny), but the repaired face images often have blurred phenomena, which can be seen in Figure 4. In some qualitative results listed, it is observed that the result of the baseline model lacks high-frequency information in the contour of the face. When the edge information is introduced into the model, the model can effectively reconstruct the overall contour of the face, but it is in the eyes, There are artifacts and distortions in details such as the mouth. Combining Table 4 and Figure 4, we can see that after introducing contour edge information and Gaussian difference detail maps, our method has obtained better quantitative indicators and qualitative results. The effectiveness of introducing different edge information is proved.

\begin{table}[]
\footnotesize
\centering 
\caption{Quantitative of different variant models.}
\resizebox{\linewidth}{!}{
\begin{tabular}{c|cccc|cc}
\hline
\multirow{2}{*}{Models} & \multicolumn{4}{c|}{FFHQ}                                                                                                             & \multicolumn{2}{c}{Widerface}                          \\ \cline{2-7} 
                        & \multicolumn{1}{c|}{FID$\downarrow$} & \multicolumn{1}{c|}{LPIPS$\downarrow$} & \multicolumn{1}{c|}{PSNR$\uparrow$} & SSIM$\uparrow$  & \multicolumn{1}{c|}{FID$\downarrow$} & NIQE$\downarrow$ \\ \hline
baseline                & \multicolumn{1}{c|}{14.88}           & \multicolumn{1}{c|}{0.218}             & \multicolumn{1}{c|}{{\ul 21.87}}    & {\ul 0.7323}    & \multicolumn{1}{c|}{15.51}           & 6.6784           \\ \hline
+Canny                  & \multicolumn{1}{c|}{12.34}           & \multicolumn{1}{c|}{{\ul 0.214}}       & \multicolumn{1}{c|}{21.72}          & 0.7270          & \multicolumn{1}{c|}{{\ul 14.81}}     & {\ul 6.4855}     \\ \hline
+Canny+DoG(Ours)              & \multicolumn{1}{c|}{\textbf{11.89}}  & \multicolumn{1}{c|}{\textbf{0.203}}    & \multicolumn{1}{c|}{\textbf{21.95}} & \textbf{0.7377} & \multicolumn{1}{c|}{\textbf{14.31}}  & \textbf{6.0863}  \\ \hline
 \hline
Ours-cycle              & \multicolumn{1}{c|}{41.62}           & \multicolumn{1}{c|}{0.245}             & \multicolumn{1}{c|}{20.52}          & 0.6548          & \multicolumn{1}{c|}{37.29}           & 6.4345           \\ \hline
Ours-w/o-cycle          & \multicolumn{1}{c|}{13.65}           & \multicolumn{1}{c|}{0.209}             & \multicolumn{1}{c|}{21.88}          & 0.7368          & \multicolumn{1}{c|}{17.45}           & 6.5511           \\ \hline

%Ours-joint              & \multicolumn{1}{c|}{18.15}           & \multicolumn{1}{c|}{0.209}             & \multicolumn{1}{c|}{21.69}          & 0.7320          & \multicolumn{1}{c|}{20.10}           & 6.2296           \\ \hline
Ours-TNet-bicubic       & \multicolumn{1}{c|}{41.01}           & \multicolumn{1}{c|}{0.223}             & \multicolumn{1}{c|}{21.83}          & 0.7295          & \multicolumn{1}{c|}{39.84}           & 8.2826           \\ \hline
Ours-SNet-bicubic       & \multicolumn{1}{c|}{20.92}           & \multicolumn{1}{c|}{\textbf{0.203}}    & \multicolumn{1}{c|}{21.58}          & 0.7319          & \multicolumn{1}{c|}{20.81}           & 6.8400           \\ \hline
Ours-TNet               & \multicolumn{1}{c|}{{\ul 15.06}}     & \multicolumn{1}{c|}{{\ul 0.212}}       & \multicolumn{1}{c|}{\textbf{22.06}} & \textbf{0.7522} & \multicolumn{1}{c|}{{\ul 18.62}}     & {\ul 6.7601}     \\ \hline
Ours-SNet(Ours)         & \multicolumn{1}{c|}{\textbf{11.89}}  & \multicolumn{1}{c|}{\textbf{0.203}}    & \multicolumn{1}{c|}{{\ul 21.95}}    & {\ul 0.7377}    & \multicolumn{1}{c|}{\textbf{14.31}}  & \textbf{6.0863}  \\ \hline

\end{tabular}
}
\label{tab:ablation}
\vspace{-4mm}
\end{table}
% Below is an example of how to insert images. Delete the ``\vspace'' line,
% uncomment the preceding line ``\centerline...'' and replace ``imageX.ps''
% with a suitable PostScript file name.
% -------------------------------------------------------------------------

%\vfill
%\pagebreak

% \begin{figure}[h]
% \begin{center}
% %\fbox{\rule{0pt}{2in} \rule{.9\linewidth}{0pt}}
%  \includegraphics[width=0.7\linewidth]{ablation.pdf}
% \end{center}
%    \caption{Qualitative results of different variant models.}
% \label{fig:ablation}
% \end{figure}

The effectiveness of each branch network: The quantitative results of all variants are shown in the lower part of Table~\ref{tab:ablation}. 'Ours-cycle' represents the TNet trained with cycle learning scheme~\cite{zhu2017unpaired}, 'Ours-w/o-cycle' represents removing DSNet. The models trained with bicubic data are recorded as 'Ours-TNet-bicubic' and 'Ours-SNet-bicubic', and our final models are recorded as 'Ours-TNet' and 'Ours-SNet'. From the results, we can observe the effectiveness of introducing the teacher-student training method and using the cycle consistency loss. Moreover, Ours-TNet obtained the highest PSNR and SSIM on the FFHQ test set.

\vspace{-4mm}
\section{CONCLUSION}
\label{sec:prior}
\vspace{-4mm}
This paper proposes a novel teacher-student FSR framework with the embedding of edge information, in which, the DeNet is used to synthesize LR data that approximates the real LR face image for the supervised training of the teacher super-resolution network. In addition, in order to facilitate the reconstruction of the overall structure and details of the face, we introduce edge information to explicitly make the network pay more attention to the recovery of high-frequency information. Extensive experiments have proved the superiority of our method and the effectiveness of all parts of the network.
%Considering the domain gap between the synthetic data and the real scene data, we only use real data to train SNet, and use the results of the teacher network as pixel-level supervision in the high-resolution space, and uses unpaired real HR images for adversarial training to make the SNet better adapt to real data.

\vspace{-4mm}
\section*{Acknowledgment}
\vspace{-2mm}
\noindent
This work is supported by the National Natural Science Foundation of China (61503277) and the National Key Research and Development Program (2019YFE0198600).

\vfill\pagebreak

\normalem
\bibliographystyle{IEEEbib}
\bibliography{strings,refs}

\end{document}